\documentclass[pdflatex,sn-mathphys-num]{sn-jnl}

\usepackage{graphicx}%
\usepackage{multirow}%
\usepackage{amsmath,amssymb,amsfonts}%
\usepackage{amsthm}%
\usepackage{mathrsfs}%
\usepackage[title]{appendix}%
\usepackage{xcolor}%
\usepackage{textcomp}%
\usepackage{manyfoot}%
\usepackage{booktabs}%
\usepackage{algorithm}%
\usepackage{algorithmicx}%
\usepackage{algpseudocode}%
\usepackage{listings}%
\usepackage{subfigure}
\usepackage{indentfirst}

\usepackage{soul}

\theoremstyle{thmstyleone}%
\theoremstyle{thmstyletwo}%

\theoremstyle{thmstylethree}%

\begin{document}


\title[Article Title]{Reliability of the Born-Oppenheimer approximation in noninteger dimensions}

\author[1,2]{\fnm{D. S.} \sur{ Rosa}}\email{derickdsr@ita.br}
\equalcont{These authors contributed equally to this work.}

\author*[1]{\fnm{T.} \sur{Frederico}}\email{tobias@ita.br}
\equalcont{These authors contributed equally to this work.}

\author[1]{\fnm{R. M.} \sur{Francisco}}\email{elfara2311@gmail.com}
\equalcont{These authors contributed equally to this work.}

\author[3]{\fnm{G.} \sur{ Krein}}\email{gastao.krein@unesp.br}
\equalcont{These authors contributed equally to this work.}

\author[3]{\fnm{M. T.} \sur{  Yamashita}}\email{marcelo.yamashita@unesp.br}
\equalcont{These authors contributed equally to this work.}

\affil[1]{ \orgname{Instituto Tecnol\'{o}gico de Aeron\'{a}utica}, \orgaddress{\street{Pr. Mal. Eduardo Gomes}, \city{S\~{a}o Jos\'{e} dos Campos}, \postcode{12228-900}, \state{SP}, \country{Brazil}}}

\affil[2]{ \orgname{Université Paris-Saclay,CNRS/IN2P3, IJCLab}, \orgaddress{\city{Orsay}, \postcode{91405},  \country{France}}}

\affil[3]{\orgdiv{Instituto de F\'isica Te\'orica}, \orgname{Universidade Estadual Paulista}, \orgaddress{\street{Rua Dr. Bento Teobaldo Ferraz, 271-Bloco II}, \city{S\~ao Paulo}, \postcode{01140-070}, \state{SP}, \country{Brazil}}}


\abstract{
We address the question of the reliability of the Born-Oppenheimer (BO) approximation for a mass-imbalanced resonant three-body system embedded in noninteger dimensions. We address this question within the problem of a system of currently experimental interest, namely $^7$Li$-^{87}$Rb$_2$. We compare the Efimov scale parameter as well as the wave functions obtained using the BO approximation with those obtained using the Bethe-Peierls boundary condition.}

\keywords{Efimov effect, Born-Oppenheimer approximation, mass-imbalance, Bethe-Peierls boundary condition, noninteger dimensions}



\maketitle

\section{Introduction}\label{sec1}

First theorized by Vitaly Efimov in 1970, the Efimov effect refers to the emergence of an infinite series of three-body bound states when the subsystems interact via short-range attractive potentials in the limit of infinite scattering lengths~\cite{EFIMOV1970563}. This phenomenon presents universal features characterized by a discrete scale invariance in the trimers spectrum, and, although originally studied in atomic and nuclear physics~\cite{Braaten:2004rn,
Frederico:2012xh,RevModPhys.89.035006}, it can be found in a wide variety of
contexts~\cite{efimovatoms,efimovdipolar,efimovphotons,
efimovpolarons}. 

The first experimental evidence of the Efimov effect was found in 2006 by Kraemer et al.~\cite{kraemer} in the study of trapped ultracold atomic systems close to Feshbach resonances~\cite{RevModPhys.82.1225} (see also~\cite{homoexp1,homoexp2,heteexp1,heteexp2,heteexp3}), while the most convincing observations for the universal Efimov scaling were provided by the heteronuclear system $^6$Li-$^{133}$Cs$_2$. The latter follows from the effect of the mass-imbalance in the Efimov three-body spectrum: it preserves the universality of Efimov states and modifies the discrete scale parameter in the three-body spectrum in a way that the attractive Efimov long-range effective interaction becomes stronger. This reduces the geometrical ratio between successive Efimov energy levels and provides a better scenario for carrying out experiments with the aim of observing multiple excited Efimov states (see, e.g.,~\cite{Naidon:2016dpf}). 

Being closely related to the study of confined cold atoms when it comes to experimental setups, it is important to investigate the influence of trap deformations on the Efimov effect. These deformations can be mimicked by performing calculations for noninteger spatial dimensions~\cite{NIELSEN2001373,mohapatra,cristensen,garridocont,induefimov,rosastm}, in which the aspect ratio of the confining trap can be related to an effective dimension $D$ using the equation: 
\begin{equation}
3(D-2)/(3-D)(D-1)= b_{ho}^2/r_{2D}^2\,,
\label{eq:Dtrap}
\end{equation} 
where  $b_{ho}$ is the oscillator length and $r_{2D}$ is the root mean square radius of the three-body system in two dimensions~\cite{garrido2024}. This relation was found and tested in Ref.~\cite{garridocont} for Gaussian and Morse short-range two-body potentials, suggesting that it is independent of the details of the two-body potential~\cite{garridoprr}. It is important to mention that, despite the advances that made it possible to compress and expand atomic clouds, creating effective two-~\cite{BEC2D} and one-dimensional~\cite{BEC1D} setups, to the best of our knowledge, continuously changing the trap geometry while controlling the Feshbach resonance is still a challenge. 

In the present work, our aim is to compare results from the study of Efimov states embedded in noninteger dimensions obtained by two different methods, namely by implementing Bethe-Peierls~\cite{bethe} (BP) boundary conditions to solve the three-body problem with contact interaction and using the Born-Oppenheimer approximation (BO). As is known, the former is used as a boundary condition for the three-body wave function when considering two-body contact interactions~\cite{PhysRevB.22.1467,Bulgac1975,EFIMOV1973157}, while the latter deals with the dynamics of the light and heavy particles by means of separated eigenvalue equations~\cite{Bellotti_2013,PhysRevA.95.062708,FONSECA1979273,rosaBO,PhysRevA.106.063305,PhysRevA.94.062707},~being its accuracy affected by the mass ratio of the particles composing the system. In this study, we observe how the results obtained by these methods are affected by variations in the noninteger dimension. To carry out our investigations, we consider the system $^7$Li$-^{87}$Rb$_2$, which is of current experimental interest.

This work is organized as follows. In section~\ref{formalism}, 
we review the two methods to solve the three-body problem with contact interactions. In Section~\ref{numres}, we discuss the numerical results. Section~\ref{sec:conclusion} presents our final considerations. 

\section{Formalism}
\label{formalism}

In this section, we review the BP approach and the BO approximation to solve the three-body problem in $D-$dimensions, which will be used to obtain and compare results for the mass-imbalanced system $^7$Li$-^{87}$Rb$_2$.  

\subsection{Bethe-Peierls Boundary Condition}
\label{sec:BP}

We start by reviewing the derivation of the $D$-dimensional three-body wave function of an Efimov state for a mass-imbalanced system at the unitary limit, according to the
approach introduced in Ref.~\cite{rosabp}. 

For a three-body system with masses $m_i$, $m_j$, $m_k$, and coordinates $\textbf{x}_{i}$, $\textbf{x}_{j}$ and $\textbf{x}_{k}$, the dynamics of the center of mass can be eliminated by using Jacobi coordinates
\begin{equation}
 \mbox{\boldmath$R$}_{i} = \textbf{x}_{j} - \textbf{x}_{k}\quad\text{and}\quad
 \mbox{\boldmath$r$}_{i} = \textbf{x}_i - \frac{m_j\textbf{x}_j+m_{k}\textbf{x}_k}{m_j + m_k} \, ,
\end{equation}
where ($i, j, k$) are taken cyclically among ($1,2,3$). The three-body wave function can be decomposed in a sum of Faddeev components that are functions of the Jacobi coordinates: 
\begin{equation}\label{eq:fullwavefunct}
\Psi(\textbf{x}_{1},\textbf{x}_{2},\textbf{x}_{3})\equiv\Psi(\mbox{\boldmath$R$},\mbox{\boldmath$r$}) = \sum_{i=1}^3
\psi^{(i)}(\mbox{\boldmath$R$}_i,\mbox{\boldmath$r$}_i)\,{\color{blue},}
\end{equation}
with each of these components satisfying the free Schrodinger equation
\begin{equation}\label{eq:schr3B}
\left[\frac{1}{2\eta_{i}}\nabla^{2}_{\mbox{\boldmath$R$}_i} +\frac{1}{2\mu_{i}} 
\nabla^{2}_{\mbox{\boldmath$r$}_i}
- E_3\right] \psi^{(i)} (\mbox{\boldmath$R$}_i,\mbox{\boldmath$r$}_i) = 0,
\end{equation}
where we have defined the reduced masses $\eta_{i} = m_{j}m_{k}/(m_{j}+m_{k})$ and $ \mu_{i} = {m_{i}(m_{j}+m_{k})}/({m_{i}+m_{j}+m_{k}})$, and $E_3$ is the three-body energy eigenvalue.

Choosing one of the pairs of Jacobi coordinates, the BP boundary condition can be applied to the total wave function, so that, in the unitary limit $a\rightarrow \infty$, we have
\begin{equation}
\label{eq:BP3B}
\hspace{-0.2cm}\left[\frac{\partial}{\partial R_i}  
R_{i}^{\frac{D-1}{2}}\Psi(\mbox{\boldmath$R$}_i,\mbox{\boldmath$r$}_i)
\right]_{R_i\rightarrow 0} = \frac{3-D}{2} 
\left[\frac{\Psi(\mbox{\boldmath$R$}_i,\mbox{\boldmath$r$}_i)}
{R_{i}^{\frac{3-D}{2}}}\right]_{R_i\rightarrow 0}.
\end{equation}
The coordinates $
 \mbox{\boldmath$R$}'_{i} = \sqrt{\eta_i}\, \mbox{\boldmath$R$}_{i}\quad$ and $\quad
 \mbox{\boldmath$r$}'_{i} = \sqrt{\mu_i}\, \mbox{\boldmath$r$}_{i}$ are introduced to simplify the form of the kinetic energies, being related to each other by the orthogonal transformations
\begin{eqnarray}
\mbox{\boldmath$R$}'_{j}& =& - \mbox{\boldmath$R$}'_{k}\cos\theta_i + \mbox{\boldmath$r$}'_{k}
\sin \theta_i, \nonumber \\
\mbox{\boldmath$r$}'_{j}& =& - \mbox{\boldmath$R$}'_{k}\sin\theta_i - 
\mbox{\boldmath$r$}'_{k}\cos \theta_i,
\end{eqnarray}
where $\tan \theta_i = \left[m_i M/(m_j\ m_k)\right]^{1/2}$, with $M = m_1 + m_2 + m_3$. 

For bosons in the s-wave state, we can define the reduced Faddeev component $
F^{(i)} (R'_{i}, r'_i) = \left( R'_{i} \ 
r'_{i}\right)^{ (D-1)/2} \psi^{(i)}(R'_i,r'_i)$, whose corresponding eigenvalue equation can be solved by using hyperspherical coordinates to separate the variables
$R'_i = \rho \sin \alpha_i$ and $r'_i = \rho \cos \alpha_i $. By doing this, we can write 
\begin{equation}
F^{(i)}(\rho,\alpha_{i})  = \mathcal{C}^{(i)} \chi(\rho)\,G^{(i)}(\alpha_{i}),
\end{equation}
\noindent where $\rho^{2}= R_i^{\prime 2}+r_i^{\prime2}$, $\alpha_i = \arctan(R'_i/r'_i)$ and the coefficients $\mathcal{C}^{(i)}$ gives the weight between the different Faddeev components for mass-imbalanced systems. 

The equations $\chi(\rho)$ and $G^{(i)}(\alpha_{i})$ satisfy the following differential equations
\begin{eqnarray}
&&\left[- \frac{\partial^{2}}{\partial \rho^{2}} + \frac{s_{n}^{2}-1/4}{\rho^{2}} + 2\kappa_0^2
\right]\chi(\rho)=0,
 \label{radialwavefunc}
\\
&&\left[- \frac{\partial^{2}}{\partial \alpha_i^{2}} -s_{n}^{2}+\frac{(D-1)(D-3)}{ \sin^2 2 
\alpha_i}\right] G^{(i)}(\alpha_i)=0, 
\label{Eq:angularD}
\end{eqnarray}
where $-\kappa_0^2 = E_3$ and $s_n$ is the Efimov parameter.
If we define $z = \cos 2\alpha_i$ and write a reduced form of the equation $G^{(i)}(z) = (1-z^2)^{1/4} g^{(i)}(z)$, Eq.~\eqref{Eq:angularD} assumes the form of the associated Legendre differential equation~\cite{legendre}, which have the known analytical solutions:
\begin{eqnarray} 
G^{(i)}(\alpha_i) &=& \sqrt{\sin2 \alpha_i} \, \Big[ P_{s_n/2-1/2}^{D/2-1}\,(\cos2\alpha_i) \nonumber \\
&-& \frac{2}{\pi}\tan\big[\pi(s_{n} -1)/2\big] Q_{s_n/2-1/2}^{D/2-1}\,(\cos2\alpha_i)\Big],\ \ \ \ \
\label{Eq:AngSol}
\end{eqnarray}
where $P_{n}^{m}(x)$ and $Q_{n}^{m}(x)$ are the associated Legendre functions. In order to have a finite value for the Faddeev component $\psi^{(i)}$ at $\rho_i =0$, $G^{(i)}(\alpha_i=\pi/2)$ must be equal to zero (recall that $r_i' = \rho \cos{\alpha_i}$). From the solution of the hyperradial
equation, Eq.~\eqref{radialwavefunc}, and the hyperangular eigenfunction, 
Eq.~\eqref{Eq:angularD}, the Faddeev components read~\cite{rosabp}:
\begin{eqnarray}
\psi^{(i)}(R'_i,r'_i) &=&\mathcal{C}^{(i)}   \frac{ K_{ s_n}\left(\sqrt{2} \kappa_0 \sqrt{  R'^{2}_{i}+  r'^{2}_{i} } 
\right) }
{ \big(  R'^{2}_{i}+ r'^{2}_{i} \big)^{D/2-1/2}}\nonumber \\
&\times&\frac{\sqrt{\sin\big[2 \arctan\left( 
R'_i/r'_i\right)\big]}}{\big\{\cos\big[ \arctan\left( R'_i/r'_i\right)\big]\ \sin\big[ \arctan\left( 
R'_i/r'_i\right)\big]\big\}^{D/2-1/2}}
\nonumber \\
&\times&\left[ P_{s_n/2-1/2}^{D/2-1}\Big\{\cos\big[2 \arctan( 
R'_i/r'_i)\big]\Big\}\right.\nonumber \\
&-&\left.\frac{2}{\pi}\tan\big[\pi(s_n-1)/2 \big] Q_{s_n/2-1/2}^{D/2-1}\Big\{\cos\big[2 
\arctan( R'_i/r'_i)\big]\Big\}\right]\, ,
\label{wavefunction}
\end{eqnarray}
\noindent where $K_{ s_n}$ is the modified Bessel function of the second kind. 

Taking the three cyclic permutations of $\{i,j,k\}$ for the BP boundary condition at the unitary limit~\cite{rosabp} with $i\neq j \neq k$, we arrive at the homogeneous linear system
\begin{eqnarray} 
&&\frac{\mathcal{C}^{(i)}}{2}
\left[ \left(\cot\alpha_i\right)^{\frac{D-1}{2}} 
\left( \sin2\alpha_i \frac{\partial}{\partial \alpha_i} 
+ D-3\right) G^{(i)} (\alpha_i) \right]_{\alpha_i\rightarrow 0}\nonumber \\
&&+ (D -2) \left[ \frac{\mathcal{C}^{(j)} \, G^{(j)}(\theta_k)}
{\left(\sin\theta_k \cos\theta_k\right)^{\frac{D-1}{2}}} 
+ \frac{\mathcal{C}^{(k)} \, G^{(k)}(\theta_j)}
{\left(\sin\theta_j \cos\theta_j\right)^{\frac{D-1}{2}}} 
\right] = 0\,,
\label{BPsystem}
\end{eqnarray} 
from where the scale parameter, $s_n$, can be obtained. For a purely imaginary scale parameter ($s_n \rightarrow is_0$), the long-range Efimov effective potential ($1/R^{2}$) in  Eq.~(\ref{radialwavefunc}) is attractive, which causes the well-known Landau ``fall-to-center", giving rise to an energy spectrum that is unbounded from below~\cite{PhysRev.47.903}. This phenomenon is closely related to the Efimov effect, with $s_n$ being named the Efimov scale parameter in this case. The Efimov regime, where $s_n$ is purely imaginary, is restricted to a range of effective dimensions that depends on the mass imbalance in the system.

\subsection{Born-Oppenheimer Approximation}
\label{sec:BO}

In what follows, we review the Born-Oppenheimer (BO) approach developed in Ref.~\cite{rosaBO}. We consider a contact potential for the light-heavy subsystem, and no interaction between the two heavy atoms. 

For studying the relative motion between the three bodies, namely two identical heavy bosons with masses $m_A$ and one light boson with mass $m_B$, we can ignore the movement of the center of mass. Considering $R$ as the distance between the two heavy atoms and $r$ the distance between the light atom and the center of mass of the heavy-heavy subsystem, the Hamiltonian in relative coordinates is given by
\begin{eqnarray} \label{eq:hamiltonian}
&&H= -\frac{\hbar^{2}}{2\eta_{B}}\nabla^{2}_{R} +V_{B}(|\textbf{R}|)+ \left(-\frac{\hbar^{2}}{2\mu_{B}}\nabla^{2}_{r}  +\sum_{j=1}^{2}V_{A}\left(\Big\vert\textbf{r}+(-1)^{j}\frac{\eta_{B}}{m_A}\textbf{R}\Big\vert\right)\right),
\end{eqnarray}
where $\eta_{B} = m_{A}/2$ and $\mu_{B} = 2m_A m_B /(2m_A +m_B )$ are the reduced masses of the system, while $V_A$ and $V_B$ denote the $AB$ and $AA$ two-body interactions, respectively. 
For $m_B \ll m_A$, and considering the lowest order of approximation in the action of the  Laplacian $\nabla_R^{2}$, we can write for the total wave function: 
\begin{eqnarray}
\Psi(\textbf{r},\textbf{R}) = \phi(\textbf{R}) \psi_R(\textbf{r})\, ,
\label{product}
\end{eqnarray}
\noindent where $\psi_R$ and $\phi$ are the wave functions of the light and heavy atoms, respectively. These conditions allow us to write separated equations for the light and heavy bosons: 
\begin{equation}
\left[-\frac{\hbar^{2}}{2\mu_{B}}\nabla^{2}_{r}  + \sum_{j=1}^{2}V_{A}\left(\Big\vert\textbf{r}+(-1)^{j}\frac{\textbf{R}}{2}\Big\vert\right)\right]\psi_R(\textbf{r}) = \epsilon(R)\psi_R(\textbf{r})\,, 
\label{eqlightatom}
\end{equation}
and
\begin{equation}
\left[- \frac{\hbar^{2}}{2\eta_{B}} \nabla_{R}^{2}+ \epsilon(R) \right] \phi(\textbf{R})
 = E_{3} \phi(\textbf{R})\,,\label{eqheavy}
\end{equation}
where $\epsilon(R)$ is the eigenvalue of the Schr\"odinger-like equation for the light-atoms, Eq.~\eqref{eqlightatom}, that enters in the heavy-heavy sub-system Schr\"odinger-like equation, Eq.~\eqref{eqheavy},
as an effective potential. $E_{3}$ is the total energy of the system,
which includes the contribution of the light atom. It is important to mention that we are considering
$D$-dimensional kinetic and potential energy operators, although not indicated.

To study the log-periodic oscillations that characterize the Efimov states, we can separate the full heavy-heavy wave function into a radial and an angular part as $\phi(\textbf{R}) = \phi(R)Y_{\textbf{L}}(\hat{R})$, where $Y_{\textbf{L}}(\hat{R})$ are the hyperspherical harmonics~\cite{HAMMER20102212}. Defining the reduced wave function $\chi(R) = R^{(D-1)/2}\phi(R)$ and considering the attractive heavy-heavy interaction generated by the presence of the light atom, we can write a radial form for the eigenvalue equation~\eqref{eqheavy} as

\begin{equation}
\left[-\frac{d^{2}}{dR^{2}}+\frac{(D-3+2L)(D-1+2L)}{4R^{2}}- \frac{m_{A}}{\hbar^{2}}\,|\epsilon(R)| \right] \chi(R)  = -\frac{m_{A}}{\hbar^{2}} E_3 \,\chi(R)
 \, ,
 \label{scaleheavy}
\end{equation}
where $L$ labels the angular momentum in integer and noninteger dimensions. In particular, for $D=3$, the centrifugal potential ${(D-3+2L)(D-1+2L)}/{4R^{2}}$ reduces to the standard centrifugal barrier. For $D=2$, $L$ is the familiar eigenvalue of the transverse component of the angular momentum operator.

In this work, we are interested in Efimov states at the unitarity limit, so that the asymptotic form of the heavy-heavy effective potential is
given by~\cite{prafrancisco}:
\begin{equation}
-\lim_{R\rightarrow 0}|\epsilon(R)|
=  -\frac{\hbar^{2}}{2\mu_{B}} \,\frac{(N-2)\,g(D)}{R^{2}},
   \label{assympsmall}
\end{equation}
where $g(D)$ comes from the solution of the transcendental 
equation 
\begin{equation}\label{eq:gD}
g(D) = \left[-\frac{ \pi \csc(D \pi/2)}
{ 2^\frac{D}{2} \Gamma({D}/{2}) K_\frac{D-2}{2}\big(\sqrt{g(D)}\big) } \right]^\frac{4}{2-D},
\end{equation}
\noindent with $K_\alpha(z)$ 
and $\Gamma(z)$ being the modified Bessel function of the second 
kind and the gamma function, respectively. When this form of potential is used in the heavy-heavy eigenvalue radial equation, Eq.~\eqref{scaleheavy}, it gives rise to the phenomenon known as Landau fall to the center, which alongside with the Thomas collapse is closely related to the Efimov effect. 

To investigate this behavior, we can write the heavy-heavy eigenvalue radial equation with a general form for the Efimov potential:
\begin{eqnarray}
 &&\left[-\frac{d^{2}}{dR^{2}}+\frac{s_{n}^2-1/4}{R^{2}}\right] \chi(R) = \frac{m_{A}}{\hbar^{2}} E_{3} \chi(R)
 \, ,
\label{deviationheavy}
\end{eqnarray}
where the strength of the potential is given by
\begin{equation}
 s_n^2 =  -\frac{m_A\ g(D)}{2\mu_{B}} + \frac{(D-2+2L)^{2}}{4}\,.
 \label{BOscale}
\end{equation}
This equation defines a critical dimension $D_c$ in which the behavior of the spectrum of the system changes from a discrete scale symmetry to a continuous one, going from a region where a imaginary solution for the scale parameter is no longer possible. To stay in the Efimov scenario, we are interested in negative values for the strength of the potential, such that $s_n \rightarrow i s_0$. In this case, $\chi(R)$ displays log-periodic oscillations, and, with the appropriated boundary condition, the eigenstates are in geometrical scaling regime. To have $s_n$ imaginary, we need the condition
\begin{equation}
  (N-2)\frac{m_A}{2\mu_{B}}  g(D)>\frac{(D-2+2L)^{2}}{4}\,. \label{eq:thomascollapse}
\end{equation}
\noindent Taking this condition into consideration for bound states, where $E_3 = -\kappa_0^{2}$, we have
\begin{equation}
  \chi(R) = \sqrt{R} K_{i s_0}(\kappa_0 R).
\end{equation}

The wave function for the light-atom at the unitarity limit was obtained in Ref.~\cite{prafrancisco}, and is written as
\begin{eqnarray}
\psi_{R}(r) &=& - \frac{R^{2-D}}{(2\pi)^\frac{D}{2}}\frac{2\mu_{B}}{\hbar^{2}}
g(D)^\frac{D-2}{4} \nonumber  \\
&\times&
\left\{ \left(\frac{r^2}{R^2}+\frac{1}{4}+\frac{r}{R}\cos(\theta_{r R}) \right)^\frac{2-D}{4} K_\frac{D-2}{2}\left( \sqrt{g(D) \left[\frac{r^2}{R^2}+\frac{1}{4}+\frac{r}{R}\cos(\theta_{r R})\right]} \right)\right. \nonumber \\
&+&\left.\left(\frac{r^2}{R^2}+\frac{1}{4}-\frac{r}{R}\cos(\theta_{r R}) \right)^\frac{2-D}{4} K_\frac{D-2}{2}\left( \sqrt{g(D) \left[\frac{r^2}{R^2}+\frac{1}{4}-\frac{r}{R}\cos(\theta_{r R})\right]} \right)\right\}.\nonumber \\
\end{eqnarray}

Now, we can write the total wave-function for our system by using Eq.~(\ref{product}), which gives
\begin{eqnarray}
    \Psi(r,R)&=& - \frac{R^{(6-3D)/2}}{(2\pi)^\frac{D}{2}}\frac{2\mu_{B}}{\hbar^{2}}
g(D)^\frac{D-2}{4} K_{is_0}(\kappa_0 R) \nonumber  \\
&\times&
\left\{ \left(\frac{r^2}{R^2}+\frac{1}{4}+\frac{r}{R}\cos(\theta_{r R}) \right)^\frac{2-D}{4} K_\frac{D-2}{2}\left( \sqrt{g(D) \left[\frac{r^2}{R^2}+\frac{1}{4}+\frac{r}{R}\cos(\theta_{r R})\right]} \right)\right. \nonumber \\
&+&\left.\left(\frac{r^2}{R^2}+\frac{1}{4}-\frac{r}{R}\cos(\theta_{r R}) \right)^\frac{2-D}{4} K_\frac{D-2}{2}\left( \sqrt{g(D) \left[\frac{r^2}{R^2}+\frac{1}{4}-\frac{r}{R}\cos(\theta_{r R})\right]} \right)\right\}.\nonumber \\
\end{eqnarray}

It is important to recall that the boundary condition $\chi(R_c)=0$ results in discrete values for $\kappa_0$, which are the solutions of $K_{is_0}\left(\kappa_0^{(n)} R_c\right)=0$. Looking for shallow bound states, where $\kappa_0^{(n)}R_c \ll1$, we have that the zeros of the Bessel function are given by
\begin{equation}
  \kappa_0^{(n)} R_c= 2e^{-\gamma}\exp{\left(-\frac{n \pi}{s_0}\right)}[1+ \mathcal{O}(s)],
\end{equation}
where $\gamma$ is the Euler constant. Taking the ratio of successive $n$-body energies gives the $D$-dimensional geometric scaling-law for Efimov states:
\begin{equation}
  \frac{E_3^{(n)}}{E_3^{(n+1)}}=\exp{\left(\frac{2\pi}{s_0}\right)}\ \ \ \ \ n\rightarrow 0,1,2,\cdots\, .
\label{energyscale}
\end{equation}

\section{Numerical Results}
\label{numres}

In this section, we present numerical results for the $^7$Li-$^{87}$Rb$_2$ system embedded in $D$ dimensions at the unitary limit. In order to contrast the exact results obtained via the BP boundary condition with those obtained by means of the BO approximation, we consider that the heavy-light subsystem interact resonantly, while the pair of heavy atoms do not interact in both approaches.

\begin{figure}[h!]
\centering
{\includegraphics[width=0.47\textwidth]{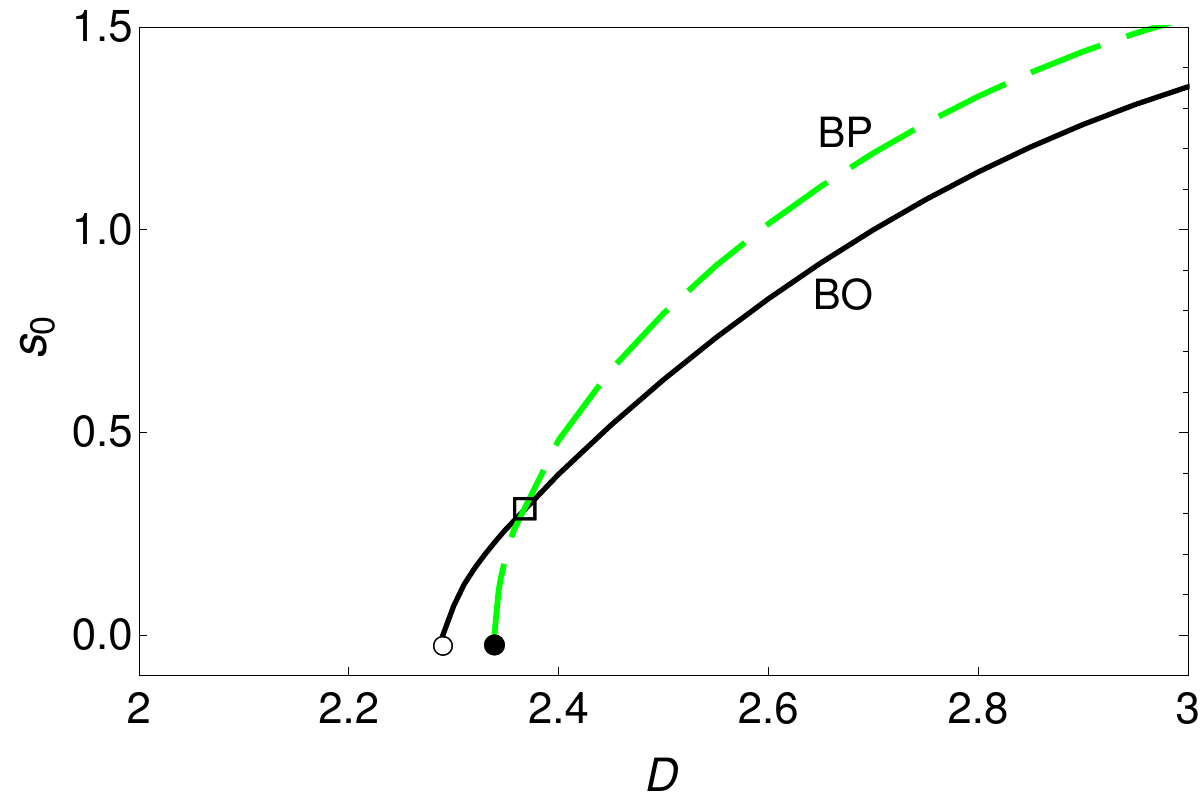}}
{\includegraphics[width=0.49\textwidth]{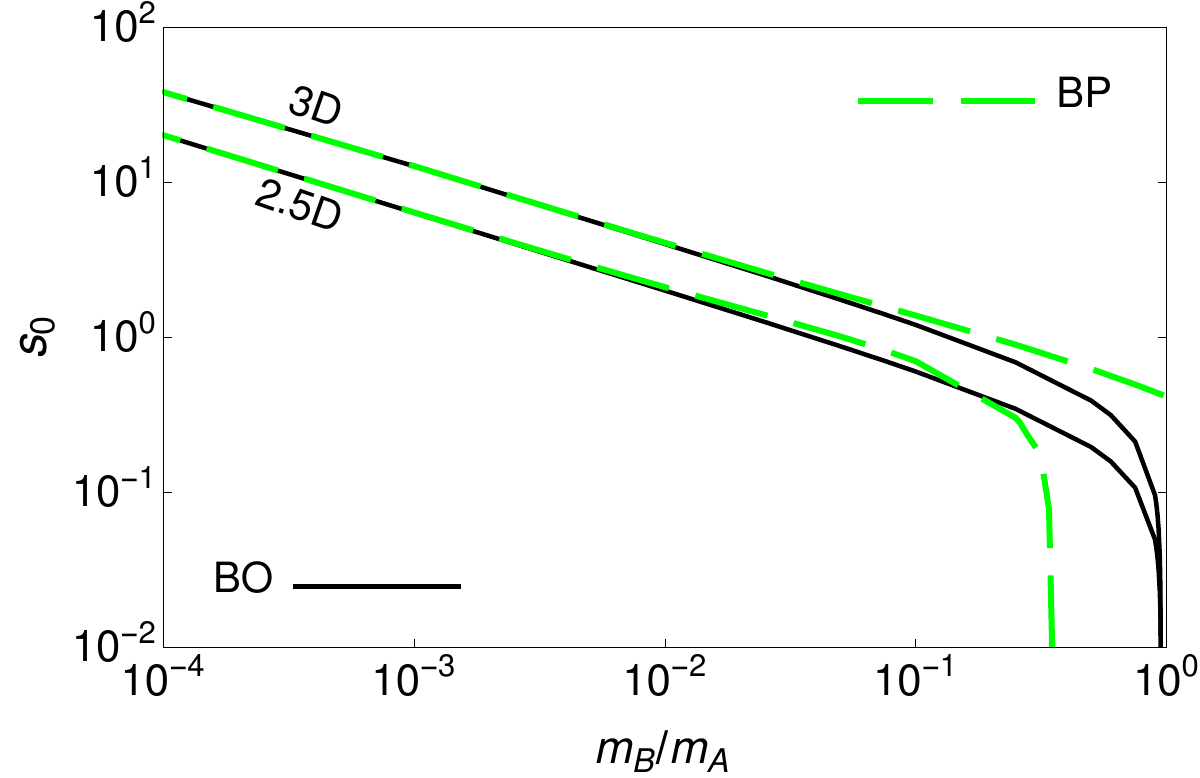}}
\caption{The Efimov discrete scale parameter for three-body molecules embedded in non-integer dimensions. Left panel: Discrete scale parameter for the $^7$Li-$^{87}$Rb$_2$ molecule in $D$ dimensions. The empty square at $D=2.365$ denotes the dimension for which the BO and BP discrete scale parameters are equal to each other. The critical values of $D$ for the BO and BP approaches are $D_{c}=2.285$ (empty circle) and $D_{c}=2.34$ (full circle), respectively. 
Right panel: Efimov scale parameter for several mass configurations of three-body system embedded in $D=3$ and $D=2.5$}. 
\label{fig1}
\end{figure}

\begin{figure}[h!]
\centering
{\includegraphics[width=0.55\textwidth]{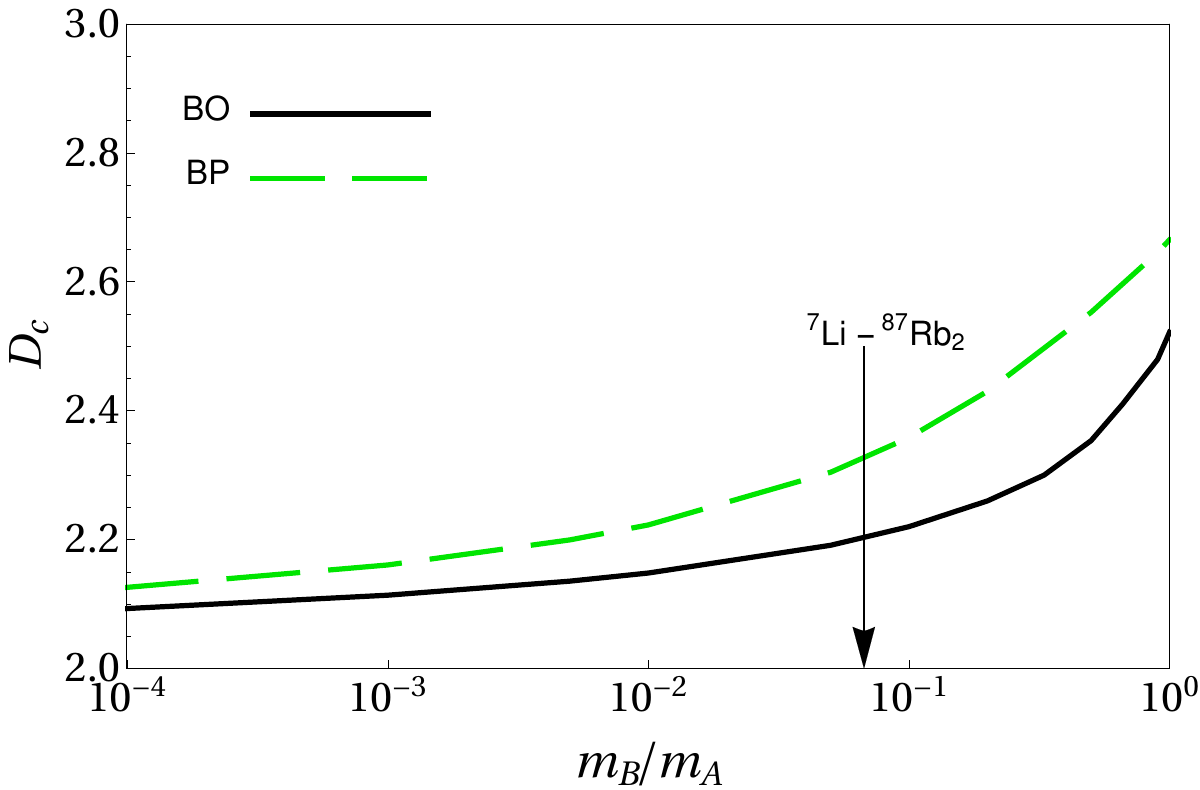}}
\caption{Critical dimension computed with the Born-Oppenheimer and Bethe-Peierls approaches considering several different mass configurations in the three-body system. }
\label{fig2}
\end{figure}

In the left panel of Fig.~\ref{fig1}, we show the Efimov scale parameter obtained via the BO approximation and via the BP boundary condition as functions of the noninteger dimension $D$ for the system $^7$Li-$^{87}$Rb$_2$. Both curves present the expected qualitative behavior, in which the scale parameter decreases as the effective dimension decreases. As is known, this behavior follows from the decrease in the strength of the Efimov long-range effective potential when the noninteger dimension decreases. In addition, starting from $D =3$, the BO values are smaller than the BP values up to $D=2.365$, denoted by the empty square, for which the BO and BP curves for the scale parameters cross each other. Beyond that point, both curves continue to decrease until they reach their respective critical dimensions $D_c$ for which $s_n = 0$.
In the BO the efimov effect is more resilient, that is, the critical dimension for this approach ($D_c= 2.285$), denoted by the empty circle, is smaller than the exact one ($D_c = 2.34$). To have a more complete picture of the comparison, in the right panel of Fig.~\ref{fig1}, the BO and BP values for the scale parameter are plotted as functions of the mass ratio $m_B/m_A$. We can see that if the mass ratio increases, the BO and the BP results will coincide at some value of the ratio. Additionally, when the mass ratio decreases, the BP goes to zero first than the BO. 

In Fig.~\ref{fig2}, we present the variation of the critical dimension $D_c$ with respect to the mass ratio $m_B/m_A$. The figure reveals that the critical dimensions obtained with the BP approach are always larger than those obtained via the BO approximation, with the difference between them decreasing as the mass ratio increases. As expected, to obtain accurate numerical results from the BO, one have to make the mass imbalance of the system extremely large. To illustrate this, for the $^7$Li$-^{87}$Rb$_2$, we have an error in the prediction of the critical dimension of $5.8$\%.

\begin{figure}[h!]
\centering
{\includegraphics[width=0.47\textwidth]{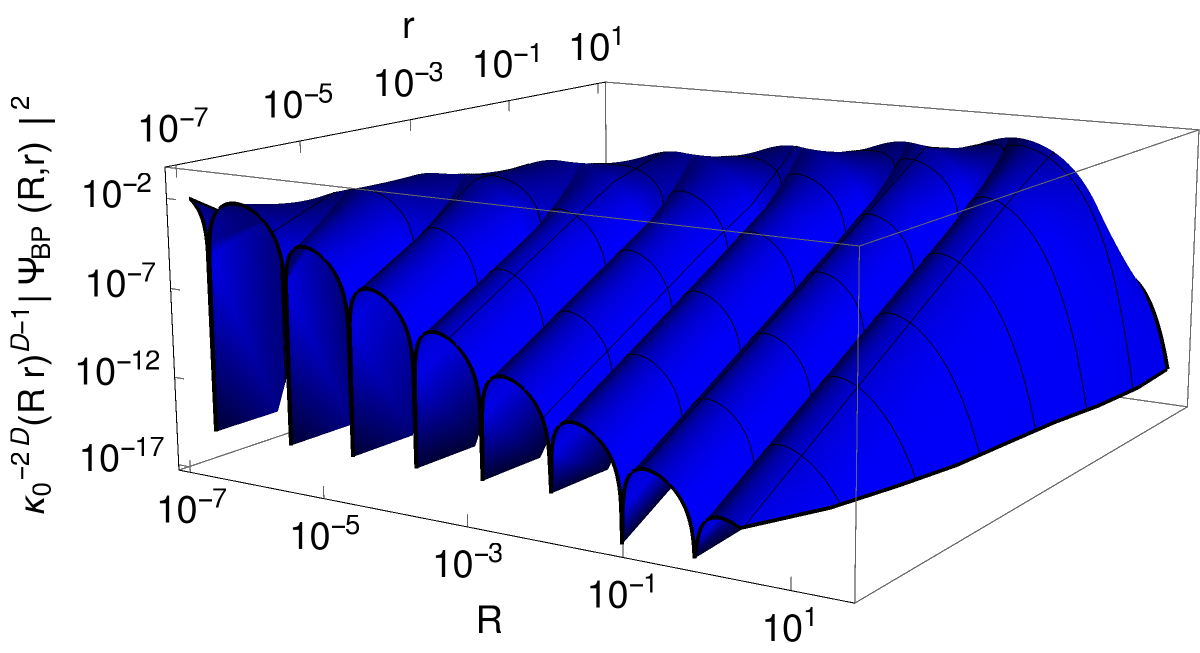}}
{\includegraphics[width=0.47\textwidth]{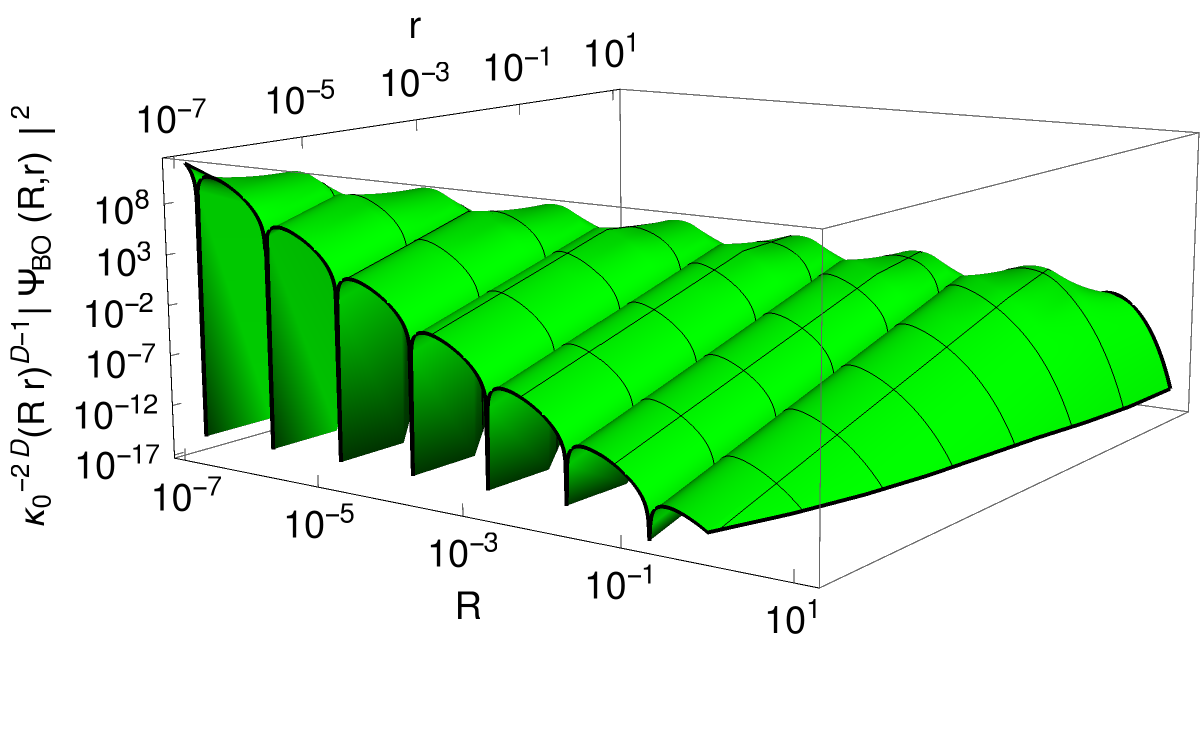}}
{\includegraphics[width=0.47\textwidth]{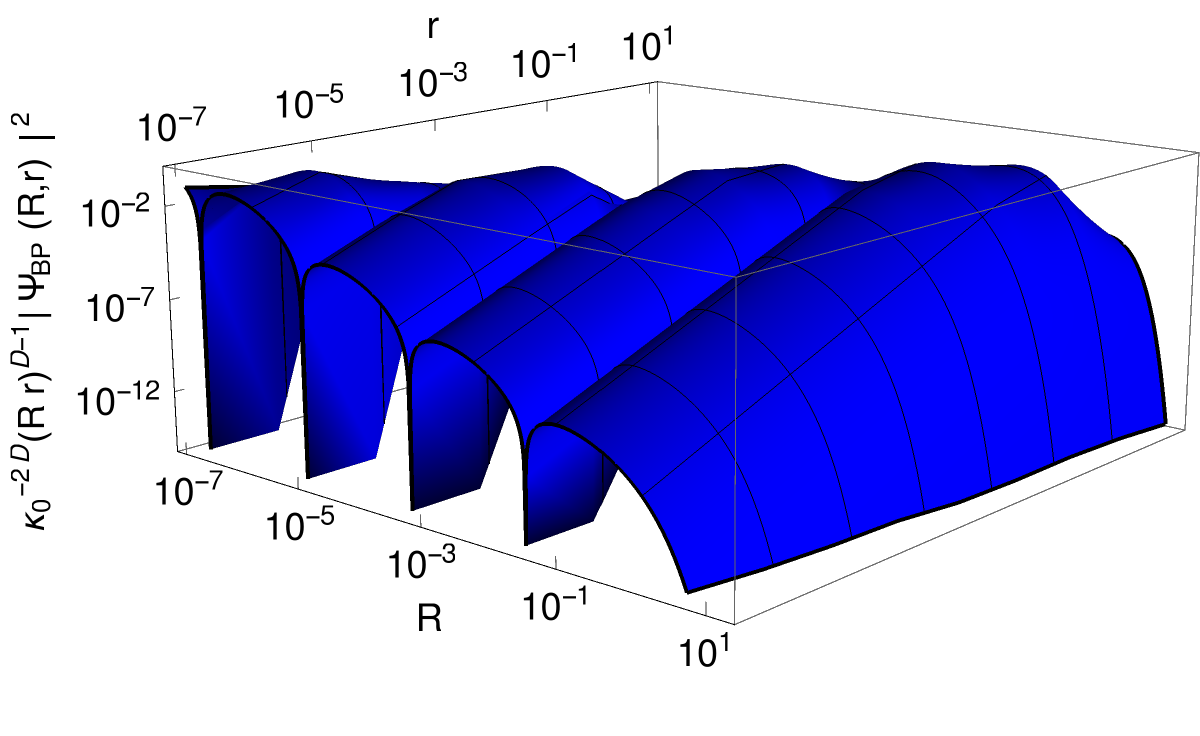}}
{\includegraphics[width=0.47\textwidth]{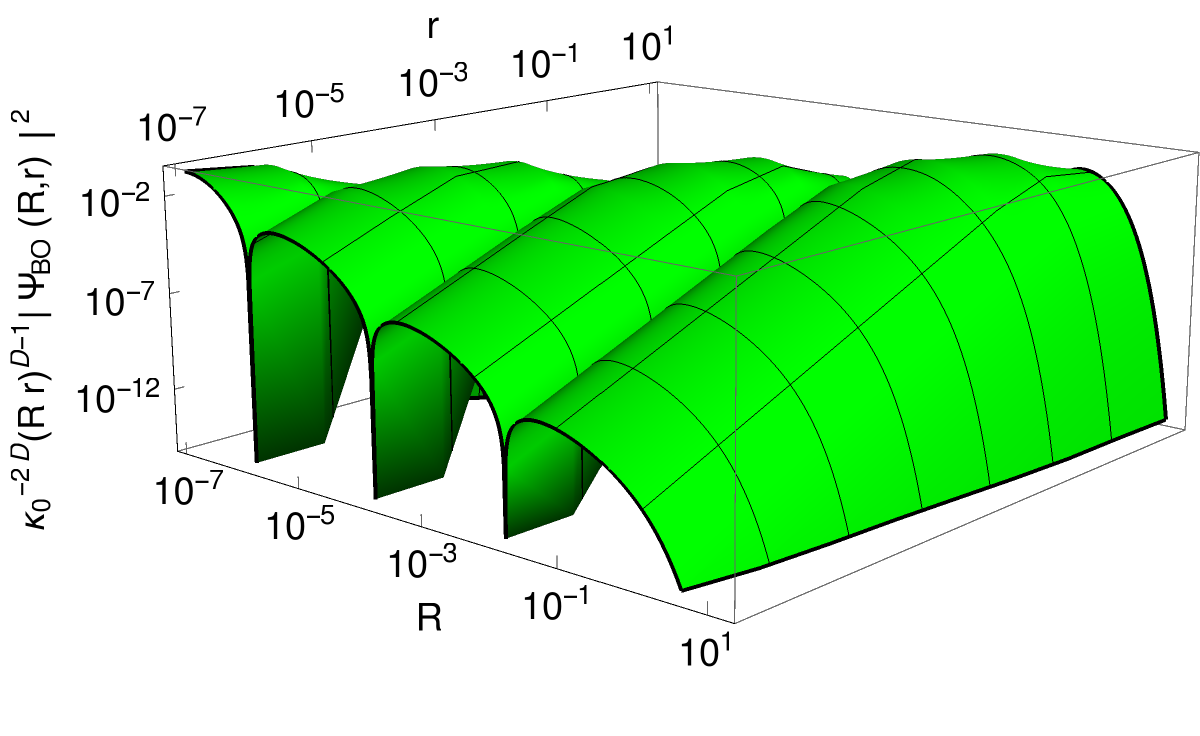}}
{\includegraphics[width=0.47\textwidth]{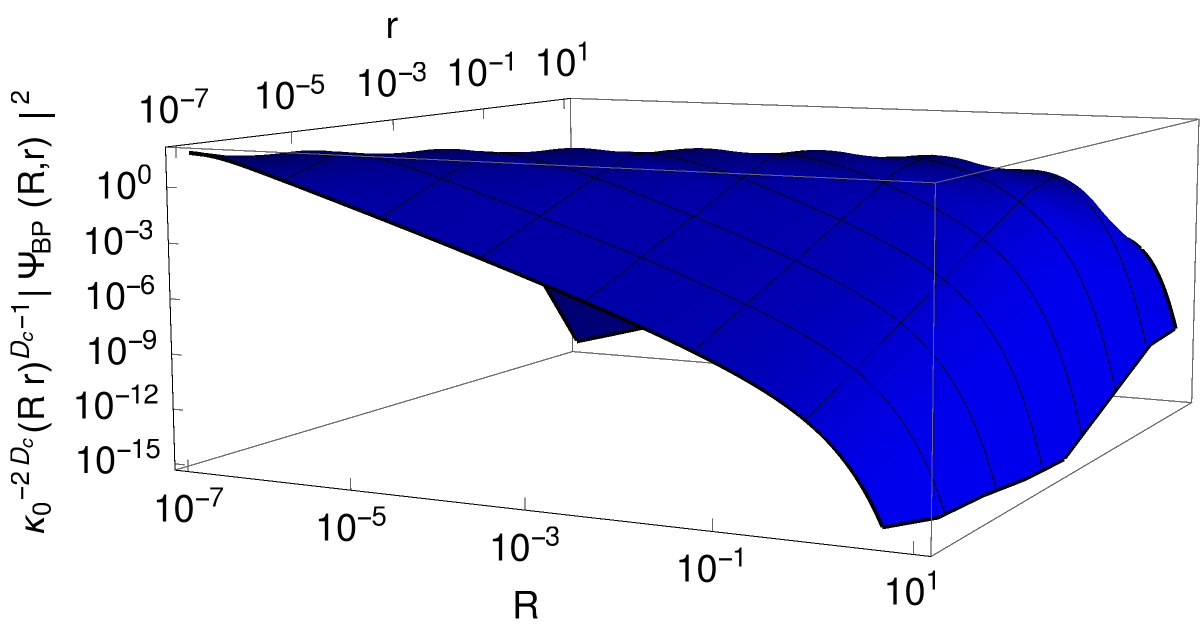}}
{\includegraphics[width=0.47\textwidth]{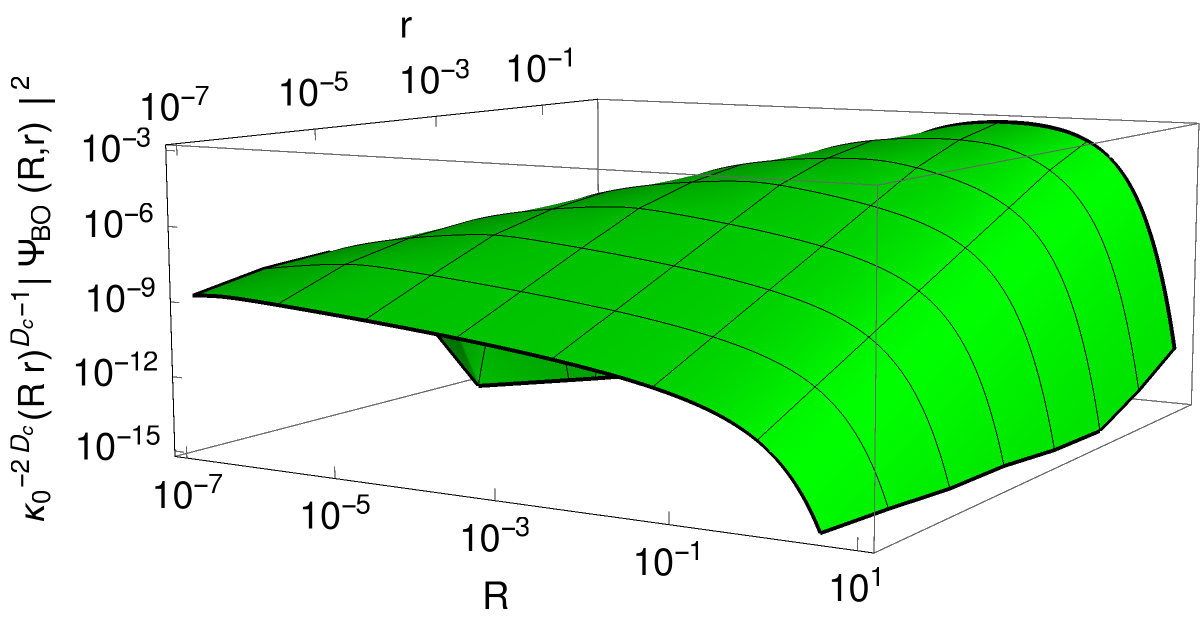}}
\caption{Dimensionless radial distributions as a function of dimensionless quantities $R = \kappa_0R_3$ (
$^{7}$Li-$^{87}$Rb relative distance) and $r=\kappa_0 r_3$ ($^{7}$Li relative distance to the $^{7}$Li-$^{87}$Rb
system). We consider the three-body system $^{87}$Rb$_2$-$^7$Li embedded in three different dimensions $3D$ (upper panels), $2.5D$ (middle panels) and at the critical dimension (lower panels). Left panels show the results from BP approach and right panels present results obtained with the BO approximation. The angle between $\vec{r}$ and $\vec{R}$ is fixed to $\pi/3$.}
\label{fig3}
\end{figure}

In Fig.~\ref{fig3}, we present the probability density $\kappa_0^{-2D}(Rr)^{D-1}\left|\Psi(R,r)\right|^2$ for the effective dimensions $D=3$ (upper panels), $D=2.5$ (middle panels) and $D=D_c$ (lower). In these plots, we can observe that for both methods, the number of nodes in the probability density diminish when the non-integer dimension decreases. In addition, we see that $\Psi_{BP}(R,r)$ has more nodes than $\Psi_{BO}(R,r)$. This can be understood by the value of the scale parameter, that, as we saw in Fig.~\ref{fig1}, is larger when obtained by BP. As known, a larger discrete scale parameter means that the Efimov excited energy states are closer to each other, causing the wave functions to concentrate more nodes in a certain region of the space for a given Efimov state. For the probability densities at the critical dimension $D_c$, shown in the lower panels of the figure, the system is no longer in the Efimov regime: there are no nodes in these distributions, and the discrete scale symmetry is no longer present.

\begin{figure}[h!]
\centering
{\includegraphics[width=0.47\textwidth]{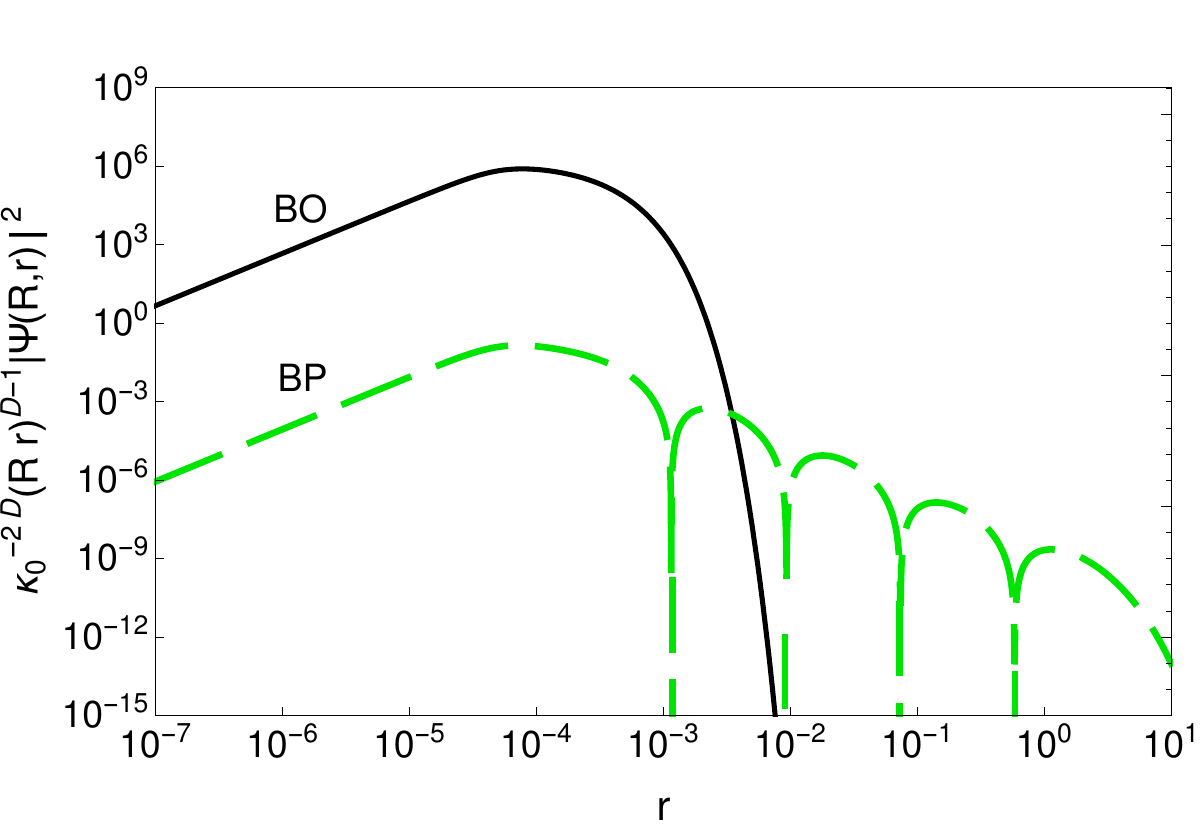}}
{\includegraphics[width=0.47\textwidth]{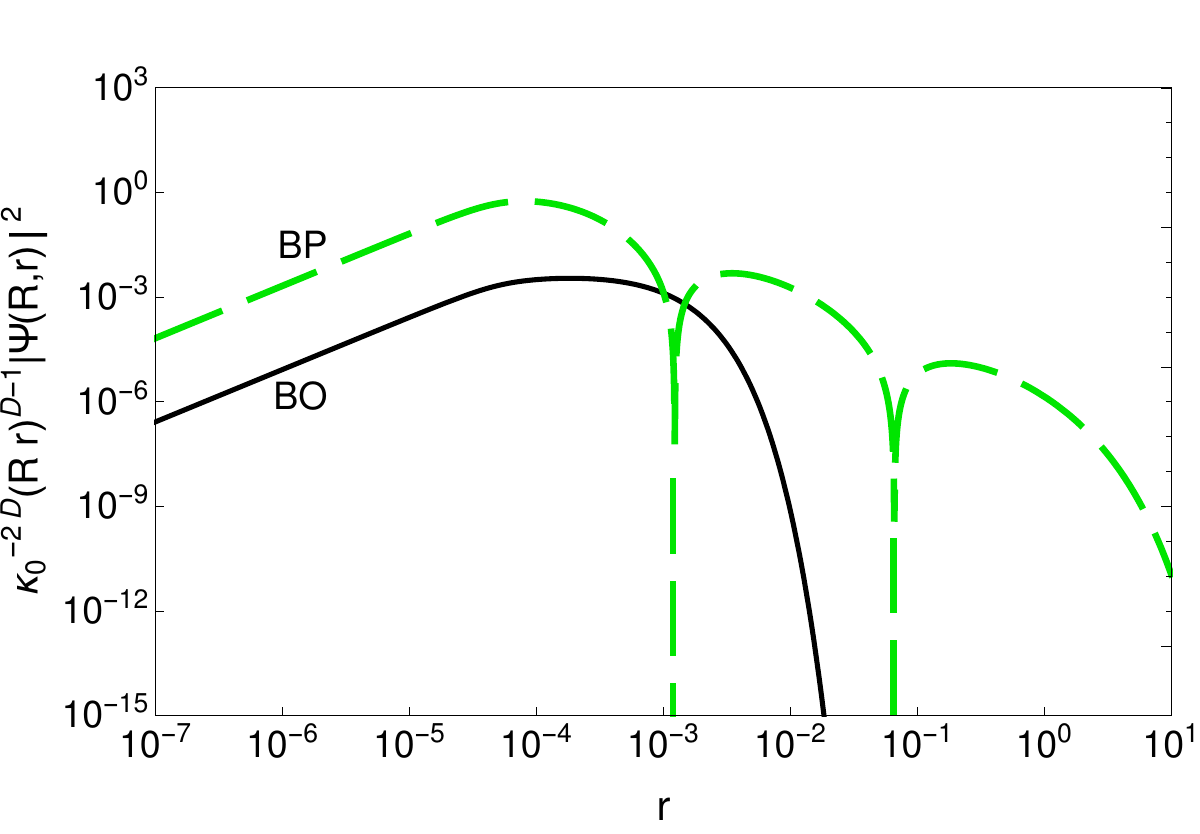}}
\caption{Dimensionless radial distribution as a function of the dimensionless quantity $r=\kappa_0 r_3$ ($r$ is the
$^{7}$Li relative distance to the $^{7}$Li-$^{87}$Rb
system) for fixed $R = \kappa_0R_3=10^{-4}$ ($R$~is~the $^{7}$Li-$^{87}$Rb relative distance). We consider the  $^{87}$Rb$_2$-$^7$Li system embedded in dimensions $D=3$ (left panel) and $D =2.5$ (right panel). The angle between $\vec{r}$ and $\vec{R}$ is fixed to $\pi/3$.}
\label{fig4}
\end{figure}

Finally, we can observe some interesting features in the BO and BP probability densities when they are plotted as functions of $\kappa_0 r$ for a fixed value $\kappa_0R=10^{-4}${\textemdash}recall that $r$ represents the distance between the light atom and the center of mass of the heavy-heavy subsystem. Fig.~\ref{fig4} displays those densities for $D=3$ (left panel) and $D=2.5$ (right panel). The graphs reveal that the BP probability density exhibits the characteristic nodes of the Efimov bound states, while the BO probability density does not exhibit them.
In order to understand this, we recall how the two methods solve the three-body problem. While in the BP boundary condition approach one takes all three atoms into consideration when looking for the long-range Efimov effective potential, in the BO approximation one freezes the degrees of freedom of the heavy atoms by assuming an adiabatic behavior, and, by doing so, one finds that at the unitary limit, the light atom generates an Efimov type potential between the heavy-heavy subsystem. This is in agreement with the picture of the Efimov long-range interaction in which a third particle is constantly "exchanged" between the other two, causing an attraction of kinetic origin. However, in the BO approximation, this interaction exists only between the two heavy atoms. Besides a quantitative discrepancy, this difference can have a significant impact on the geometry of the system, which we intend to investigate in a future work.

\section{Conclusion}\label{sec:conclusion}

In this work, we investigated the reliability of the Born-Oppenheimer approximation when applied in the study of Efimov states embedded in effective noninteger dimensions. To accomplish this, we considered mass-imbalanced systems composed of two heavy atoms and a light one and compared the numerical results obtained using the Born-Oppenheimer appromation with those obtained using the Bethe-Peierls boundary condition approach. In order to have compatible results, in both methods we considered a resonant contact interaction only between the light and heavy atoms, and no interaction in the heavy-heavy subsystem.

Considering the trimer $^7$Li$-^{87}$Rb$_2$, we found that the qualitative behaviors obtained through BO and BP are in agreement, with the scale parameter that characterizes the discrete scale symmetry between Efimov states going to zero when the noninteger dimension is decreased. In turn, the quantitative results exhibit a discrepancy that can be observed in the values of the scale parameters: the values obtained through the BP approach are larger than those obtained with the BO approximation until $D=2.365$. At this point, the situation is inverted and remains in this way until the critical dimension $D_c$, where the discrete scale symmetry is broken. Going further in this investigation, we observed that the critical dimensions ($D_c$) for the BP results are smaller than the BO ones, becoming equivalent in the limit of extreme mass imbalance. We found for the system studied that the discrepancy for the critical dimensions is $5.8$\% with respect to the mass ratio $m_B/m_A$. 

We have also constructed dimensionless radial distributions for the system in three different noninteger dimensions, namely $D=3$, $D=2.5$, and $D=D_c$. We observed that, for all these $D$ values, the BP distributions have more nodes than the BO ones, a behavior that follows from the larger values of scale parameters obtained with the BP approach. Moreover, the number of nodes in the radial distributions decreases when the noninteger dimension becomes smaller. At the critical dimension $D_c$, for which the discrete scale symmetry is no longer present, there are no more nodes in the distributions.

In summaray, as expected, the BO approximation exhibits results that are in qualitative agreement with those obtained via the more accurate BP method. However, the numerical results show a considerable discrepancy, with the discrepancy becoming negligible only in the case of extreme mass imbalance.

\section*{Acknowledgments}

This work was partially supported by Funda\c{c}\~{a}o de Amparo \`{a} Pesquisa do Estado de S\~{a}o Paulo (FAPESP) [grant nos. 2017/05660-0 and 2019/07767-1 (T.F.), 2023/08600-9 (R.M.F.), 2023/02261-8 (D.S.R.) and 2018/25225-9 (G.K.)] and Conselho Nacional de Desenvolvimento Cient\'{i}fico e Tecnol\'{o}gico (CNPq) [grant nos. 
306834/2022-7 (T.F.), 302105/2022-0 (M.T.Y.), and  309262/2019-4 (G.K.)].





\bibliography{sn-bibliography}

\end{document}